\documentstyle[sprocl,epsfig]{article}

\def\be{\begin{equation}}
\def\ee{\end{equation}}
\def\bea{\begin{eqnarray}}
\def\eea{\end{eqnarray}}

\begin{document}

\title{THE LITTLE HIGGS BOSON AT A PHOTON COLLIDER\\}

\author{ HEATHER E.\ LOGAN }

\address{Department of Physics, University of Wisconsin,\\
1150 University Avenue, Madison, Wisconsin 53706 USA}


\maketitle\abstracts{
We calculate the partial widths of the light Higgs boson in the Littlest
Higgs model.  The loop-induced Higgs coupling to photon pairs, which is
especially sensitive to effects of the new TeV-scale particles running in
the loop, can be probed with high precision at a photon collider in the
process $\gamma \gamma \to H \to b \bar b$.  Using the parameters of the
Littlest Higgs model measured at the LHC one can calculate a prediction
for this process, which will serve as a test of the model and as a probe
for a strongly-coupled UV completion at the 10 TeV scale.
}
  

At a photon collider, the Higgs boson can be produced in the 
$s$-channel via $\gamma\gamma \to H$.  This allows
a high-precision measurement of the $\gamma\gamma H$ coupling,
which is limited by systematic uncertainties at the LHC and by
statistics at an $e^+e^-$ linear collider.  Numerous studies 
[1]
indicate that the rate for $\gamma\gamma \to H \to b \bar b$
can be measured to about 2\% for a SM-like Higgs boson with 
115 GeV $\leq M_H \leq 140$ GeV.  Other Higgs decay modes will 
be measured with lower precision.

The $\gamma\gamma H$ coupling comes from the dimension-6 operator
\begin{equation}
  {\mathcal L} = \frac{C}{\Lambda^2} H^{\dagger} H F_{\mu\nu} F^{\mu\nu},
\end{equation}
where $H$ is the Higgs doublet, $F^{\mu\nu}$ is the electromagnetic 
field strength tensor, $\Lambda$ is the mass scale that characterizes
the interaction, and $C$ is a dimensionless coefficient.
In the SM, this operator is induced by $W$ and $t$ loops.  Taking
$C = e^2/16\pi^2$ for a loop-suppressed electromagnetic interaction
yields $\Lambda = 170$ GeV for the SM -- the right scale for the 
$W$ and $t$ loops.

A 2\% measurement of $\gamma\gamma\to H$ allows a probe of new physics.
For weakly-coupled new physics
($C_{new} = e^2/16\pi^2$), scales up to 1.2 TeV (0.74 TeV) can be probed
at the 95\% confidence ($5\sigma$) level.
For strongly-coupled new physics ($C_{new} = 1$), scales up
to 48 TeV (31 TeV) can be probed at the 95\% confidence ($5\sigma$) level.
In this talk, based on 
[2],
we apply the $\gamma\gamma \to H$ measurement to the
Littlest Higgs model in order to probe for a strongly coupled UV 
completion at the 10 TeV scale.

\begin{figure}[ht] 
     \begin{center}
     \resizebox{\textwidth}{!}
     {\rotatebox{270}{\includegraphics{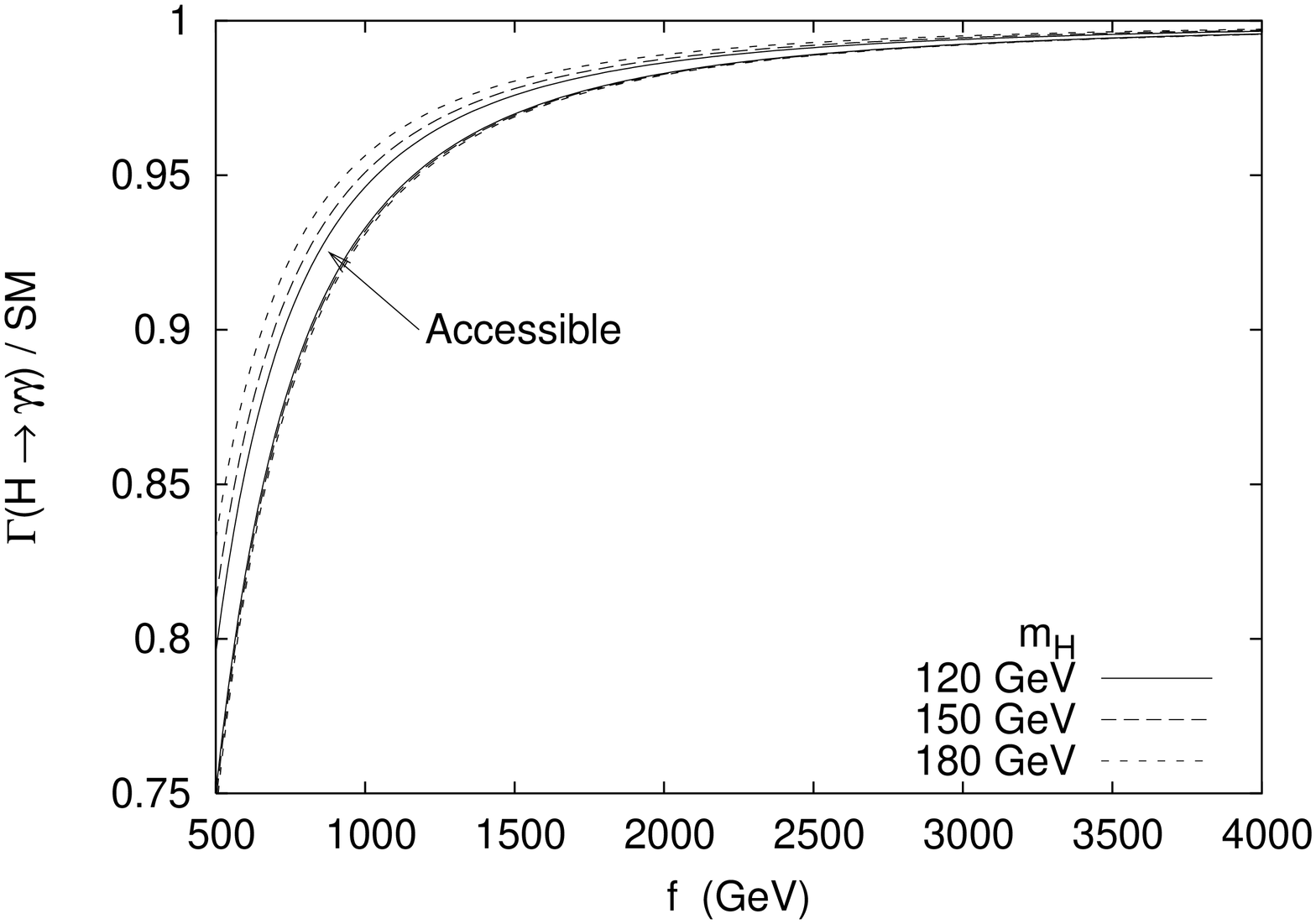}}
      \rotatebox{270}{\includegraphics[50,50][555,600]{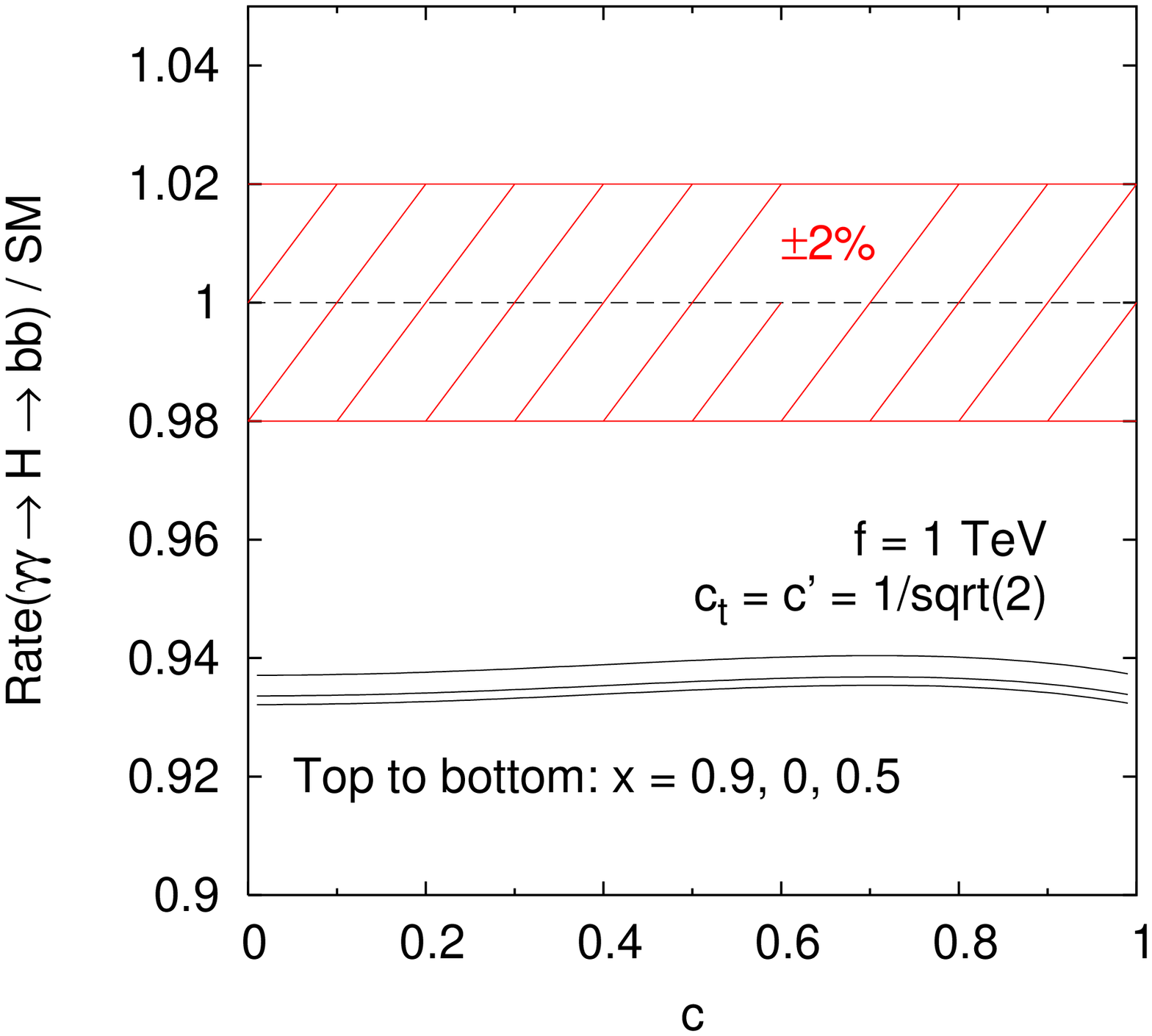}}}
     \end{center}
     \caption{(Left) Range of $\Gamma(H \to \gamma\gamma)$
accessible in the Littlest Higgs model as a function of $f$, normalized
to its SM value, for $M_H = 120$, 150 and 180 GeV.
(Right) Rate for $\gamma\gamma \to H \to b \bar b$, normalized to
its SM value, as a function of $c$ for $x = 0$, 0.5 and 0.9 (solid lines),
with $f = 1$ TeV, $c_t = c^{\prime} = 1/\sqrt{2}$,
and $M_H = 115$ GeV.} 
     \label{fig:gg}
     \end{figure}

The Littlest Higgs model 
[3]
stabilizes the little hierarchy
between the weak scale and the 10 TeV scale by making the SM Higgs doublet
a pseudo-Nambu-Goldstone boson.
At the TeV scale the model contains an SU(2) triplet of gauge 
bosons $Z_H,W_H$, a heavy isosinglet quark $T$, a scalar isotriplet
$\Phi$, and a U(1) gauge boson $A_H$ (an alternate version
contains no $A_H$).
The model is parameterized by an overall scale $f \sim$ TeV, mixing 
angles $\cos\theta \equiv c$ and $\cos\theta^{\prime} \equiv c^{\prime}$
in the extended gauge sector (these can be traded for $M_{Z_H}$ and 
$M_{A_H}$ once $f$ is known), a ratio of Yukawa couplings $c_t$ in the 
extended top sector, and a Higgs sector parameter $x$ proportional to
the isotriplet vacuum expectation value
[4,5].
The corrections to $\Gamma(\gamma\gamma \to H)$ 
in the Littlest Higgs model are shown in Fig.~\ref{fig:gg}
(left)
[5].
The corrections to the rest of the Higgs decay partial widths,
calculated in 
[2],
are all ${\mathcal O}(v^2/f^2)$
and are roughly the same size in each Higgs decay channel.  We thus focus
on the best-measured channel;
the correction to the rate for $\gamma\gamma \to H \to b \bar b$ is shown
in Fig.~\ref{fig:gg} (right).

\begin{figure}[ht] 
     \begin{center}
     \resizebox{0.9\textwidth}{!}
     {\rotatebox{270}{\includegraphics[50,50][555,600]{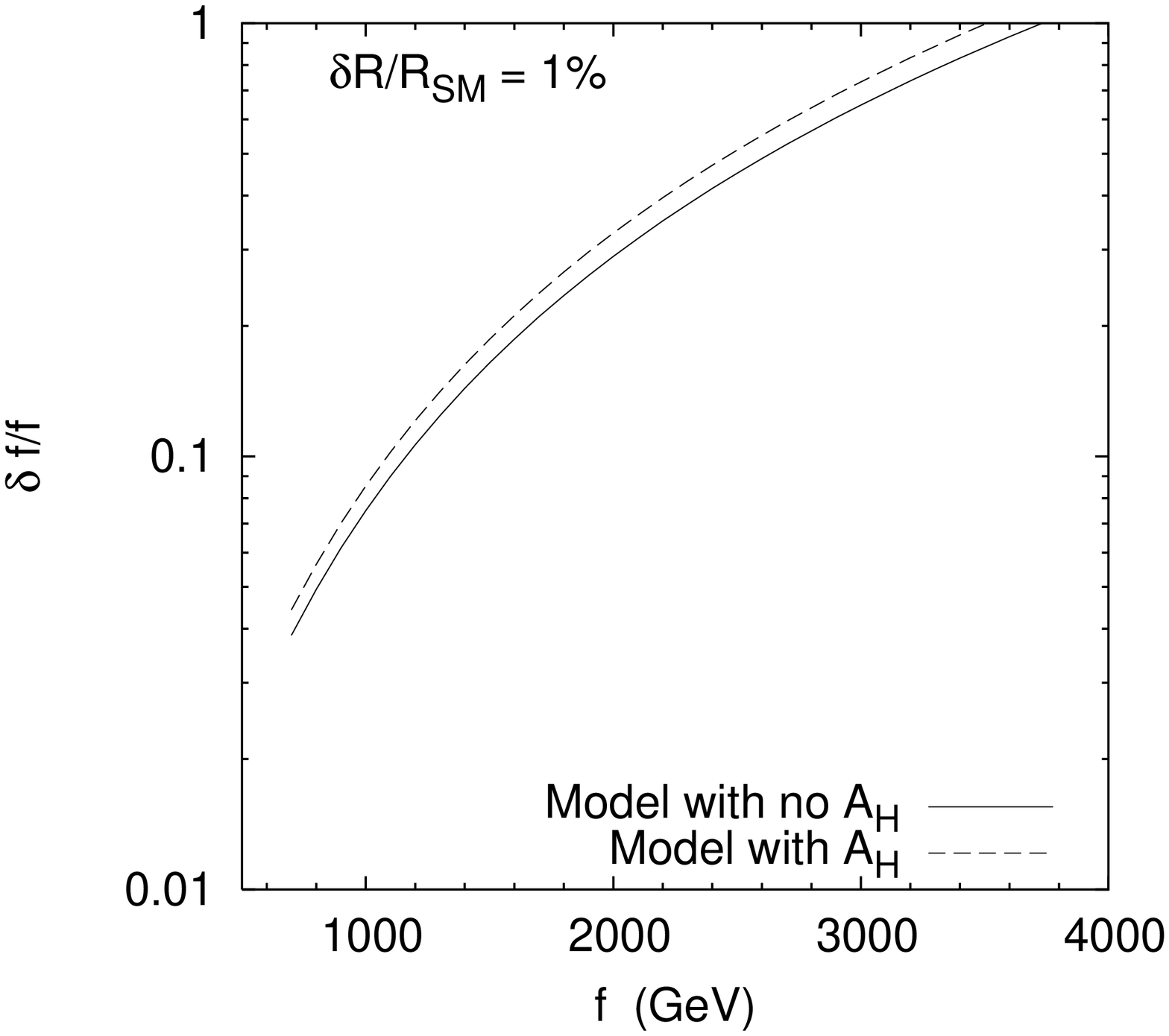}}
      \rotatebox{270}{\includegraphics[50,50][555,600]{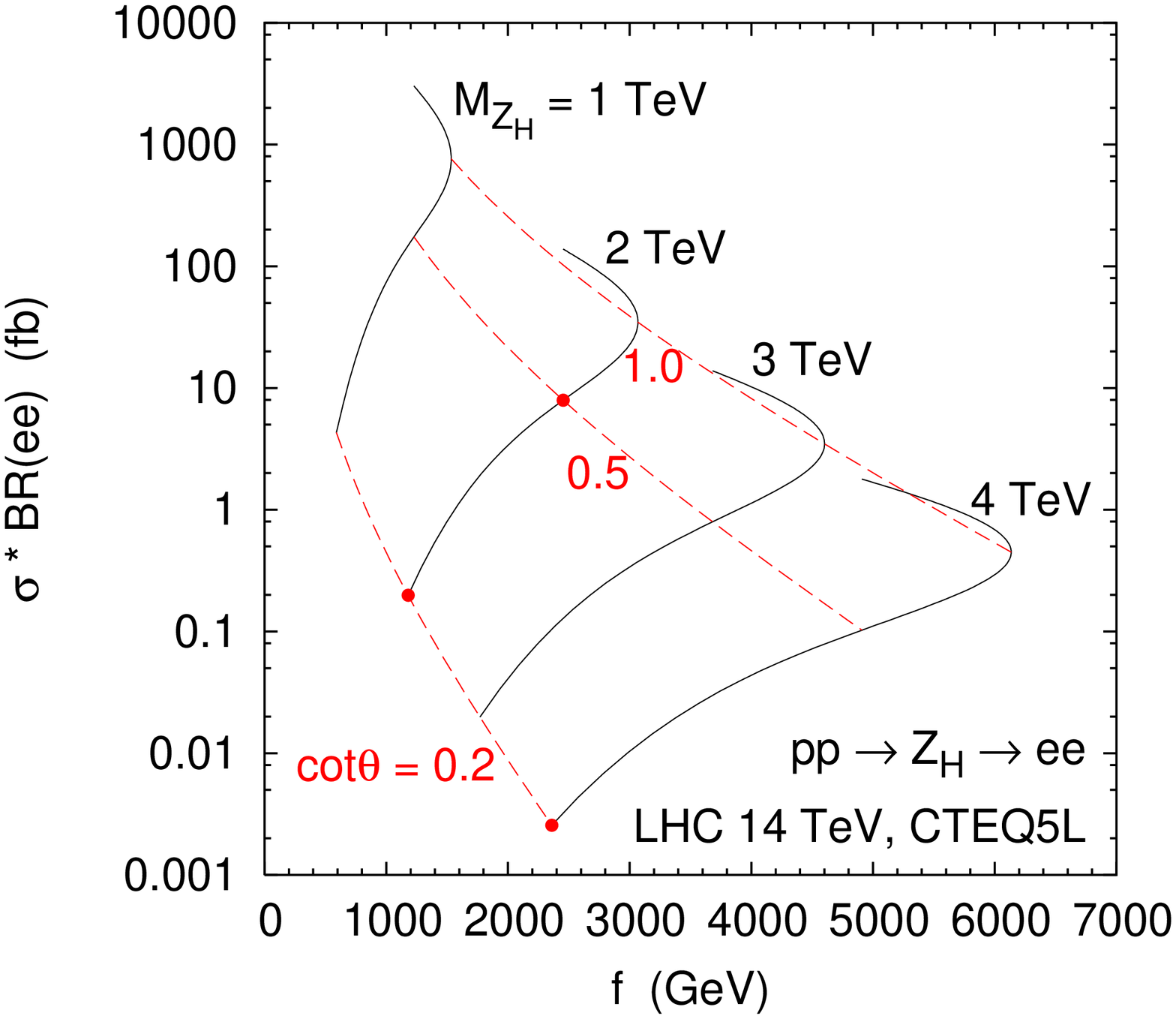}}}
     \end{center}
     \caption{(Left) Precision on $f$ required for a 1\%
parametric uncertainty for the Littlest Higgs model with (dashed) 
and without (solid) an $A_H$ particle.
(Right) Cross section times branching ratio for $Z_H$ into dielectrons
at the LHC as a function of $f$.  Solid black lines are contours of 
constant $M_{Z_H}$ and dashed red lines are contours of constant 
$\cot\theta$.} 
     \label{fig:f}
     \end{figure}

A 2\% measurement of the $\gamma\gamma\to H \to b \bar b$ rate can be used
both to test the Littlest Higgs model by probing the effects of the new 
weakly-coupled particles at 1--3 TeV and to search for a strongly-coupled
UV completion of the model at a few tens of TeV.
To do this, one must measure the model parameters (at the LHC) well enough
to be able to predict the rate for $\gamma\gamma\to H \to b \bar b$ with
sufficient precision; we take 1\% parametric uncertainty as our standard.
The sensitivity of $\gamma\gamma\to H \to b \bar b$ to $c_t$ and $M_{A_H}$
is very weak; these parameters need not be measured.  The sensitivity
to $x$ is also rather weak; $x$ need only be measured 
(from, e.g., its effect on the $W$ mass)
if $f$ is low $\sim 1$ TeV and $x$ is close to 1.
$f$ must be measured
with a precision of several percent to several tens of percent,
see Fig.~\ref{fig:f} (left).  Combining $M_{Z_H}$ with the rate for
$Z_H \to e^+e^-$ at the LHC allows $f$ to be
extracted, see Fig.~\ref{fig:f} (right).

In conclusion,
a future photon collider should be able to measure the rate for
$\gamma\gamma\to H \to b \bar b$ with 2\% precision for 
$115 \leq M_H \leq 140$ GeV.  At the same time,
the rate for this process in the Littlest Higgs model
can be reliably calculated over a large range of model parameter space
based on LHC measurements of the parameters.
Comparing prediction with measurement
then allows a probe of the UV completion of the Littlest Higgs model
at 10 TeV.
A strongly coupled UV completion contributes to $\gamma\gamma\to H$ at
the same order as the TeV-scale particles, giving a several percent
correction for $f \sim 1$--3 TeV.
A weakly coupled UV completion should not affect $\gamma\gamma\to H$ 
at an observable level; in this case, the measurement provides a test
of the model's consistency.

\end{document}